\documentclass[a4paper,12pt]{article}
\usepackage{latexsym}
\newcommand{\be}{\begin{equation}}
\newcommand{\ee}{\end{equation}}
\begin{document}


\begin{titlepage}

\begin{center}
{\Large {\bf A comment on BCC crystalization in higher dimensions }}\\\vspace{1in} {\large S. Elitzur \footnote{elitzur@vms.huji.ac.il}, B. Karni \footnote{boaz.karni@mail.huji.ac.il}, E. Rabinovici \footnote{eliezer@vms.huji.ac.il}}\\ \vspace{0.5in}
{{\it Racah Institute of Physics, The Hebrew University, \\ Jerusalem 91904, Israel}}\\
\end{center}

\begin{abstract}
The result  that near the melting point three-dimensional crystals have an octahedronic structure is generalized to higher flat non compact dimensions.
\end{abstract}
\end{titlepage}

\tableofcontents

\section{Introduction}
In their paper "Should All Crystals Be bcc? Landau Theory of Solidification and Crystal Nucleation" \cite{[1]} , S. Alexander and J. McTague review the Landau theory of solidification \cite{[2]} which predicts that solids in three space dimensions form a bcc(body centered cubic) structure and pointed out that indeed under certain circumstances many solids near their melting point exhibit this behavior.  Landau's theory of phase-transitions implies that an octahedron structure in momentum space gives a global extremum to the free-energy, and thus is the preferred structure. This is true in three spatial dimensions. In two dimensions either triangular or  honeycombed lattices are predicted to form depending on the sign of the appropriate order parameter. Solidification  occurs when an inhomogeneous configuration  becomes energetically favored compared to a homogeneous one. The condensing density waves allow various patterns of spontaneously broken translational and rotational invariance. For a review of somewhat more complex spontaneous breaking of space time symmetries see i.e.  \cite{[3]}.

 In string theory higher dimensional systems are natural, in fact ten dimensional  supersymmetrical systems have a special role as being stable.
Assuming the microscopic theory allows the formation of solids in higher dimensions it is interesting to investigate if  general structures can emerge generically.
In tachyonic string backgrounds closed and open tachyon fields are also forced to have a non-zero wave number, condensing this field leads to structures similar to that of solidification, however not necessarily in three dimensions \cite{[4]}. This motivates generalizing the study of Alexander and McTague to higher dimensions, which is the subject of this paper.  Following their work we assume the structure of regular polytops and find that the result remains true in higher flat non compact dimension.

In section 2 we review the work of S. Alexander and J. McTague and in section 3 we generalize it to more than three space dimensions.

\section{Review}
In Landau's theory of phase transition, one expands the free-energy, \be\Phi ={{\Phi }_{2}}+{{\Phi }_{3}}+...\label{eq1}\ee Where ${\Phi }_{n}$ is of order n in the density field $\rho$.\\
The presence of a ${\Phi }_{3}$ term  is essential for the solidification to occur.  A ${\Phi }_{4}$ term provides stabilization.

The second-order term, in momentum-space, is:
\be{{\Phi }_{2}}=\int{{{d}^{3}}{{q}_{1}}{{d}^{3}}{{q}_{2}}A\left( \overrightarrow{{{q}_{1}}},\overrightarrow{{{q}_{2}}} \right)\delta \left( \overrightarrow{{{q}_{1}}}+\overrightarrow{{{q}_{2}}} \right){{\rho }_{\overrightarrow{{{q}_{1}}}}}{{\rho }_{\overrightarrow{{{q}_{2}}}}}}=\int{{{d}^{3}}qA\left( {\vec{q}} \right){{\rho }_{\overrightarrow{q}}}{{\rho }_{-\overrightarrow{q}}}}\ee
With ${{\rho }_{{\vec{q}}}}$ being the Fourier component of the density function, i.e. \\ $\rho \left( {\vec{r}} \right)=\frac{1}{{\left(2\pi\right)}^{3/2}}\int{{{d}^{3}}q{{\rho }_{{\vec{q}}}}{{e}^{-i\vec{q}\cdot \vec{r}}}}$. The form of the function $A\left( {\vec{q}} \right)$ depends on the microscopic theory. The delta-function ensures translational-invariance and that $A\left( {\vec{q}} \right)$ is a function of a single $\vec{q}$. Moreover  isotropy implies that  $A\left( {\vec{q}} \right)$ depends actually only on the magnitude of $\vec{q}$. Translational invariance will be spontaneously broken for those class of functions $A\left( {\vec{q}} \right)$ for which the magnitude $Q$ of the vector $\vec{q}$ does not vanish at the extremum. In such a  ground state the configuration  ${{\rho }_{q}}$ gets support only on the momentum space sphere of this radius. The second order term is then proportional to $\int{d{{\Omega }_{{\vec{Q}}}}{{\rho }_{{\vec{Q}}}}{{\rho }_{\vec{-Q}}}}$, which equals to $\int{{{d}^{3}}r{{\rho }^{2}}}(r)$, henceforth denoted by ${{\rho }^{2}}$.The magnitude of ${{\rho }^{2}}$ is fixed by the fourth order term.

 In momentum-space the third-order term is:
\be{{\Phi }_{3}}=\int{{{d}^{3}}{{q}_{1}}{{d}^{3}}{{q}_{2}}{{d}^{3}}{{q}_{3}}B\left( \overrightarrow{{{q}_{1}}},\overrightarrow{{{q}_{2}}},\overrightarrow{{{q}_{3}}},T \right)\delta \left( \overrightarrow{{{q}_{1}}}+\overrightarrow{{{q}_{2}}}+\overrightarrow{{{q}_{3}}} \right){{\rho }_{\overrightarrow{{{q}_{1}}}}}{{\rho }_{\overrightarrow{{{q}_{2}}}}}{{\rho }_{\overrightarrow{{{q}_{3}}}}}}\ee Here T stand for all of the thermodynamical quantities on which B depends. Since  the magnitude of the $\overrightarrow{{{q}_{i}}}'s$ is already fixed, one can re-write this term as:

\be{{\Phi }_{3}}={{B}_{Q}}\left( T \right)\int{d{{\Omega }_{1}}d{{\Omega }_{2}}d{{\Omega }_{3}}\delta \left( \overrightarrow{{{Q}_{1}}}+\overrightarrow{{{Q}_{2}}}+\overrightarrow{{{Q}_{3}}} \right){{\rho }_{\overrightarrow{{{Q}_{1}}}}}{{\rho }_{\overrightarrow{{{Q}_{2}}}}}{{\rho }_{\overrightarrow{{{Q}_{3}}}}}}\ee  One sees that ${{\Phi }_{3}}$ gets a contribution only from configurations in which the $\overrightarrow{{{Q}_{i}}}'s$ form equilateral triangles. Thus rotational invariance is also spontaneously broken. Since the position-space density ($\rho \left( {\vec{r}} \right)$) is real, each configuration has to satisfy the equation ${{\rho }_{-\overrightarrow{{{Q}_{i}}}}}={{\rho }_{\overrightarrow{{{Q}_{i}}}}}^{\dagger }$, which gives us \be\rho\left({\vec{r}}\right)=\frac{1}{\sqrt{2}}\sum\limits_{i=1}^{n}{\left({{\rho}_{\overrightarrow{{{Q}_{i}}}}}{{e}^{i{{{\vec{Q}}}_{i}}\vec{r}}}+{{\rho }_{\overrightarrow{-{{Q}_{i}}}}}{{e}^{-i{{{\vec{Q}}}_{i}}\vec{r}}} \right)}\ee as possible density configurations.  For such a configuration, with 2n values of q for which ${{\rho }_{q}}\ne 0$, (if q is such a value, then necessarily -q is also such a value),  in order to maximize ${{\Phi }_{3}}$, at a given ${{\rho }^{2}}$, all ${{\rho }_{\overrightarrow{{{Q}_{i}}}}}'s$ should be of the same magnitude.  One obtains that  $n{{\left| {{\rho }_{\overrightarrow{{{Q}_{i}}}}} \right|}^{2}}={{\rho }^{2}}$, so the term ${{\rho }_{\overrightarrow{{{Q}_{1}}}}}{{\rho }_{\overrightarrow{{{Q}_{2}}}}}{{\rho }_{\overrightarrow{{{Q}_{3}}}}}$ is proportional to ${{n}^{-{3/2}}}$.
One can  use this result to determine the configuration of the crystal. The momenta ${{Q}_{i}}$ for which ${{\rho }_{\overrightarrow{{{Q}_{i}}}}}\ne 0$ form some polyhedron in momentum-space. This polyhedron has exactly $2n$ edges. The ${{\Phi }_{3}}$ term gets a contribution from each face of that polyhedron which is an equilateral triangle. ${{\Phi }_{3}}$ is then proportional to ${{\left( \frac{{{N}_{e}}}{2} \right)}^{-3/2}}\cdot {{N}_{t}}$, where ${{N}_{e}}$ is the number of edges of the polyhedron, and ${{N}_{t}}$ is the number of faces that are equilateral triangles.
The regular, convex, three-dimensional polyhedron that maximizes the ${{\Phi }_{3}}$ term, is the octahedron \cite{[1]}. Although two tetrahedra give the same result,  this is essentially the same configuration, i.e. the momentum vectors that generate one generate the other.

\section{Higher-Dimensional Space}
In higher dimensions, the same ideas apply. One considers the regular, convex polytopes in higher dimensions.
Postponing for a moment the discussion of the case of a four-dimensional space, which is a bit more complicated, we discuss five or more dimensions. It is known \cite{[5]} that in five dimensions or more there are only three regular polytopes, all convex: The n-simplex, the n-hypercube, and its dual the n-cross polytope, which is the n-dimensional analog of the octahedron. The free energy corresponding to each polytope configuration depends on the number of elements in its one-skeleton (edges), and on the number of elements of its two-skeleton (two-dimesional faces) that are equilateral triangles. The hypercube does not contribute in this case, since its two-skeleton does not include any triangles, equilateral or otherwise, so the value of ${{\Phi }_{3}}$ for it is zero.
For the n-simplex the number of elements in its one-skeleton is given by ${n+1 \choose 2}$ \cite{[5]} and the number of elements in its two-skeleton is given by ${n+1 \choose 3}$ \cite{[5]}. All the elements of the two-skeleton are equilateral triangles, so we should consider all of them in our calculation. One should notice, though, that a single n-simplex does not contain any pair of opposite edges, so two n-simplexes have to be considered, which gives ${{N}_{e}}=2\cdot {n+1 \choose 2}$ elements in the one-skeleton and ${{N}_{t}}=2\cdot{n+1 \choose 3}$ elements in the two-skeleton. The ${{\Phi }_{3}}$ term is then proportional to  \be\left(\frac{N_e}{2}\right)^{-3/2}\cdot N_t={n+1 \choose 2}^{-3/2}\cdot 2 \cdot {n+1 \choose 3}=\frac{2^{5/2}}{6}\frac{n-1}{\left(n\left(n+1\right)\right)^{1/2}}\ee
Turning our attention now to the cross-polytope, we have ${{2}^{2}}\cdot{n \choose 2}$ elements in the one-skeleton \cite{[5]}, and ${{2}^{3}}\cdot {n \choose 3}$ elements in the two-skeleton \cite{[5]}, all of which are equilateral triangles. This will give a value of \be\left(\frac{N_e}{2}\right)^{-3/2}\cdot N_t=\left(2 \cdot {n \choose 2}\right)^{-3/2}\cdot 2^3\cdot {n\choose 3}=\frac{4(n-2)}{3\sqrt{n(n-1)}} \label{2skel} \ee
Dividing these two results, we get that the ratio between the value of ${{\Phi }_{3}}$ for the cross-polytope and the value of ${{\Phi }_{3}}$ for the simplex is \be\sqrt{2} \cdot \frac{n-2}{n-1}\cdot{\left(\frac{ n+1 }{n-1} \right)}^{1/2}\ee which is greater than one for n greater than three.
So we see that for five dimensions, or more, the cross-polytope, which is the n-dimensional analogue of the octahedron, is the preferred momentum-space configuration in these dimensions as well.
In the special case of four dimensions one has six convex regular polytopes \cite{[5]}. A direct check shows that here also the cross-polytope, called a 16-cell in four dimensions, is preferred. Thus the result holds for any dimension greater than two.
Since (\ref{2skel}) is a rising function of $n$ for $n>2$, a higher dimensional polytope will always be preferable to a lower dimensional one, ruling out liquid-crystals as a possible structure for a high-dimensional crystal, for the case studied here of a scalar order parameters. Tensorial order parameters allow more structure already at $n=3$.
To determine the structure of the resulting lattice in position-space, notice that the vertices of the n cross polytope are of the form ${{e}_{i}}$, $1\le i\le n$, with $\pm 1$ in the i'th coordinate, and 0 in all other. This means that the edges of this polytope are n-vectors, with $\pm 1$ in the i'th and j'th place, $i\ne j$, and 0 in all others, i.e. a vector of the form  \be\left( 0,0,...,\underbrace{\pm 1}_{\scriptsize \textrm{i'th place}}\normalsize,0,0,...,\underbrace{\pm 1}_{\scriptsize \textrm{j'th place}}\normalsize,0,...,0 \right)\ee Thus, the lattice in momentum-space is generated by n linearly-independent such vectors. The reciprocal lattice, in position-space, is the set of vectors where scalar products with the momentum lattice vectors are integers. It is generated by vectors with $\pm \frac{1}{2}$ in all their entries. This is exactly a b.c.c. lattice.

\section{Conclusions}
We have shown that the result  that near melting a bcc lattice is preferred, is true in any number of flat non compact dimensions.\\

Several issues require  further study. One is  how to generalize this analysis to compact dimensions of various topologies. The global structure of the manifold will further constrain the allowed configurations requiring in some cases to deal with commensurability aspects.  Another is the case of configurations that form polytopes that are not convex or not regular, as may occur on the sphere. It is possible that such a configuration, which would be preferable to the cross-polytope, exists. Such a configuration will not form a crystal, but it is possible, if it has sufficient symmetry, that it will form a quasi-crystal.

\section*{Acknowledgments}
We thank David Mukamel for many discussions.\\
The work of E. Rabinovici is partially supported by the Humbodlt foundation and the American-Israeli Bi-National Science Foundation.\\
The work of S. Elitzur and E. Rabinovici is partially supported by a DIP grant H, 52, the Einstein Center at the Hebrew University,  and the Israel Science Foundation Center of Excellence.

\end{document}